\documentclass[11pt]{article}

\usepackage{times}
\usepackage{latexsym}
\usepackage{amsfonts,amsthm,amssymb}
\usepackage{euscript}
\usepackage{amstext}
\usepackage{graphicx}
\usepackage{color}
\usepackage{fullpage}
\usepackage{tikz}

\newtheorem{lemma}{Lemma}[section]
\newtheorem{theorem}[lemma]{Theorem}

\newtheorem{remark}[lemma]{Remark}

        {\hspace*{\fill}$\Box$\par}

\newcommand{\lemref}[1]{Lemma~\ref{lemma:#1}}

\newcommand{\lemlab}[1]{\label{lemma:#1}}

\newcommand{\figlab}[1]{\label{fig:#1}}

\newcommand{\opt}{\textrm{\sc OPT}}
\newcommand{\etal}{et al.\ }
\newcommand{\eps}{\epsilon}
\newcommand{\st}{S} 
\newcommand{\Algorithm}[1]{{\texttt{\bf{#1}}}} 
\newcommand{\sbg}{\Algorithm{SSF-W}} 
\newcommand{\sug}{\Algorithm{SSF}} 
\newcommand{\lf}{\Algorithm{LF}}

\newcommand{\lwf}{\Algorithm{LWF}}
\newcommand{\fifo}{\Algorithm{FIFO}}
\newcommand{\partially}{\mathcal{A}} 
\newcommand{\zeroly}{\mathcal{B}}      
\newcommand{\bwf}{\Algorithm{BWF-W}} 
\newcommand{\srfw}{\Algorithm{SRF-W}} 
\begin{document}

\title{Minimizing Maximum Response Time and \\ Delay Factor in Broadcast Scheduling}

\author{
Chandra Chekuri\thanks{Department of Computer Science, University of Illinois, 201 N.\ Goodwin Ave., Urbana, IL 61801.
{\tt chekuri@cs.uiuc.edu}. Partially supported by NSF grants CCF-0728782
 and CNS-0721899. }
 \and Sungjin Im\thanks{Department of Computer Science, University of
Illinois, 201 N.\ Goodwin Ave., Urbana, IL 61801. {\tt im3@uiuc.edu}} \and Benjamin Moseley\thanks{Department of
Computer Science, University of Illinois, 201 N.\ Goodwin Ave., Urbana, IL 61801. {\tt bmosele2@uiuc.edu}. Partially
supported by NSF grant CNS-0721899. } }

\begin{titlepage}
\def\thepage{}

\maketitle

\begin{abstract}
  We consider {\em online} algorithms for pull-based broadcast
  scheduling. In this setting there are $n$ pages of information at a
  server and requests for pages arrive online. When the server serves
  (broadcasts) a page $p$, all outstanding requests for that page are
  satisfied. We study two related metrics, namely maximum response
  time (waiting time) and maximum delay-factor and their {\em
    weighted} versions. We obtain the following results in the
  worst-case online competitive model.
  \begin{itemize}
  \item We show that FIFO (first-in first-out) is $2$-competitive even
    when the page sizes are different. Previously this was known only
    for unit-sized pages \cite{ChangEGK08} via a delicate argument.
    Our proof differs from \cite{ChangEGK08} and is perhaps more
    intuitive.
  \item We give an online algorithm for maximum delay-factor
    that is $O(1/\eps^2)$-competitive with $(1+\eps)$-speed for unit-sized
    pages and with $(2+\eps)$-speed for different sized pages. This
    improves on the algorithm in \cite{ChekuriM09} which required
    $(2+\eps)$-speed and $(4+\eps)$-speed respectively. In addition
    we show that the algorithm and analysis can be extended to
    obtain the same results for maximum {\em weighted} response time
    and delay factor.
  \item We show that a natural greedy algorithm modeled after LWF
    (Longest-Wait-First) is not $O(1)$-competitive for maximum delay
    factor with any constant speed even in the setting of standard
    scheduling with unit-sized jobs. This complements our upper bound
    and demonstrates the importance of the tradeoff made in our
    algorithm.
  \end{itemize}
\end{abstract}

\end{titlepage}

\section{Introduction}
We consider \emph{online} algorithms in pull-based broadcasting. In
this model there are $n$ pages (representing some form of useful
information) available at a server and clients request a page that
they are interested in. When the server transmits a page $p$, all
outstanding requests for that page $p$ are satisfied since it is
assumed that all clients can simultaneously receive the
information. It is in this respect that broadcast scheduling differs
crucially from standard scheduling where each jobs needs its own
service from the server. We distinguish two cases: all the pages are
of same size (unit-size without loss of generality) and when the pages
can be of different size. Broadcast scheduling is motivated by several
applications in wireless and LAN based systems \cite{AcharyaFZ95,
  AksoyF98, Wong88}.  It has seen a substantial interest in the
algorithmic scheduling literature starting with the work of Bartal and
Muthukrishanan \cite{BartalM00}; see \cite{KalyanasundaramPV00}.  In
addition to the applications, broadcast scheduling has sustained
interest due to the significant technical challenges that basic
problems in this setting have posed for algorithm design and
analysis. To distinguish broadcast scheduling from ``standard'' job
scheduling, we refer to the latter as unicast scheduling --- we use
requests in the context of broadcast and jobs in the context of
unicast scheduling.

In this paper, we focus on scheduling to minimize two related
objectives: the maximum response time and the maximum delay factor. We
also consider their {\em weighted} versions.  Interestingly, the
maximum response time metric was studied in the (short) paper of
Bartal and Muthukrishnan \cite{BartalM00} where they claimed that the
online algorithm FIFO (for First In First Out) is $2$-competitive for
broadcast scheduling, and moreover that no deterministic online
algorithm is $(2-\eps)$-competitive. (It is easy to see that FIFO is
optimal in unicast scheduling). Despite the claim, no proof was
published. It is only recently, almost a decade later, that Chang et
al. \cite{ChangEGK08} gave formal proofs for these claims for
unit-sizes pages. This simple problem illustrates the difficulty of
broadcast scheduling: the ability to satisfy multiple requests for a
page $p$ with a single transmission makes it difficult to relate the
total ``work'' that the online algorithm and the offline adversary
do. The upper bound proof for FIFO in \cite{ChangEGK08} is short but
delicate. In fact, \cite{BartalM00} claimed $2$-competitiveness for
FIFO even when pages have different sizes. As noted in previous work
\cite{BartalM00,EdmondsP03,PruhsU03}, when pages have different sizes,
one needs to carefully define how a request for a page $p$ gets
satisfied if it arrives midway during the transmission of the page. In
this paper we consider the sequential model \cite{EdmondsP03}, the
most restrictive one, in which the server broadcasts each page
sequentially and a client receives the page sequentially without
buffering; see \cite{PruhsU03} on the relationship between different
models. The claim in \cite{BartalM00} regarding FIFO for different
pages is in a less restrictive model in which clients can buffer and
take advantage of partial transmissions and the server is allowed to
preempt. The FIFO analysis in \cite{ChangEGK08} for unit-sized pages
does not appear to generalize for different page sizes. Our first
contribution in this paper is the following.

\begin{theorem}
\label{thm:fifo}
  FIFO is $2$-competitive for minimizing maximum response time in
  broadcast scheduling even with different page sizes.
\end{theorem}

Note that FIFO, whenever the server is free, picks the page $p$ with
the earliest request and {\em non-preemptively} broadcasts it.  Our
bound matches the lower bound shown even for unit-sized pages, thus
closing one aspect of the problem. Our proof differs from that of
Chang et al.; it does not explicitly use the unit-size assumption and
this is what enables the generalization to different page sizes.  The
analysis is inspired by our previous work on maximum delay factor
\cite{ChekuriM09} which we discuss next.

\smallskip
\noindent
{\bf Maximum (Weighted) Delay Factor and Weighted Response Time:}
The delay factor of a schedule is a metric recently introduced in
\cite{ChangEGK08} (and implicitly in \cite{BenderCT08}) when requests
have deadlines. Delay factor captures how much a request is delayed
compared to its deadline. More formally, let $J_{p,i}$ denote the
$i$'th request of page $p$. Each request $J_{p,i}$ arrives at
$a_{p,i}$ and has a deadline $d_{p,i}$. The finish time $f_{p,i}$ of a
request $J_{p,i}$ is defined to be the earliest time after $a_{p,i}$
when the page $p$ is sequentially transmitted by the scheduler
starting from the beginning of the page. Note that multiple requests
for the same page can have the same finish time. Formally, the delay
factor of the job $J_{p,i}$ is defined as $ \max\{1, \frac{f_{p,i} -
  a_{p,i}}{d_{p,i} - a_{p,i}}\}$; we refer to the quantity $S_{p,i} =
d_{p,i} - a_{p,i}$ as the {\em slack} of $J_{p,i}$. For a more detailed motivation of delay factor, see
\cite{ChekuriM09}. Note that for unit-sized pages, delay factor generalizes response time since one could set $d_{p,i}
= a_{p,i} + 1$ for each request $J_{p,i}$ in which case its delay factor equals its response time. In this paper we are
interested in online algorithms that minimize the maximum delay factor, in other words the objective function is $\min
\max_{p,i} \{1, \frac{f_{p,i} - a_{p,i}}{d_{p,i} - a_{p,i}}\}$. We also consider a related metric, namely {\em
weighted} response time. Let $w_{p,i}$ be a non-negative weight associated with $J_{p,i}$; the weighted response time
is then $w_{p,i} (f_{p,i} - a_{p,i})$ and the goal is to minimize the maximum weighted response time. Delay factor and
weighted response time have syntactic similarity if we ignore the $1$ term in the definition of delay factor --- one
can think of the weight as the inverse of the slack. Although the metrics are some what similar we note that there is
no direct way to reduce one to the other. On the other hand, we observe that upper bounds for one appear to translate
to the other. We also consider the problem of minimizing the maximum weighted delay factor $\min \max_{p,i} w_{p,i}
\{1, \frac{f_{p,i} - a_{p,i}}{d_{p,i} - a_{p,i}}\}$.

Surprisingly, the maximum weighted response time metric appears to not
have been studied formally even in classical unicast scheduling;
however a special case, namely maximum {\em stretch} has received
attention. The stretch of a job is its response time divided by its
processing time; essentially the weight of a job is the inverse of its
processing time.  Bender \etal \cite{BenderCM98,BenderMR02}, motivated
by applications to web-server scheduling, studied maximum stretch and
showed very strong lower bounds in the online setting. Using similar
ideas, in some previous work \cite{ChekuriM09}, we showed strong
lower bounds for minimizing maximum delay factor even for unit-time
jobs. In \cite{ChekuriM09}, constant competitive algorithms were given
for minimizing maximum delay factor in both unicast and broadcast
scheduling; the algorithms are based on resource augmentation
\cite{KalyanasundaramP95} wherein the algorithm is given a speed $s >
1$ server while the offline adversary is given a speed $1$
server. They showed that $\sug$ (shortest slack first) is
$O(1/\eps)$-competitive with $(1+\eps)$-speed in unicast
scheduling. $\sug$ does not work well in the broadcast scheduling. A
different algorithm that involves waiting, $\sbg$ (shortest slack
first with waiting) was developed and analyzed in \cite{ChekuriM09};
the algorithm is $O(1/\eps^2)$-competitive for unit-size pages with
$(2+\eps)$-speed and with $(4+\eps)$-speed for different sized
pages.
In this paper we obtain improved results by altering the analysis of
$\sbg$ in a subtle and important way. In addition we show that the
algorithm and analysis can be altered in an easy fashion to obtain the
same bounds for weighted response time and delay factor.

\begin{theorem}
  \label{thm:delayfactor}
  There is an algorithm that is $(1+\eps)$-speed
  $O(1/\eps^2)$-competitive for minimizing maximum delay factor in
  broadcast scheduling with unit-sized pages.  For different sized
  pages there is a $(2+\eps)$-speed $O(1/\eps^2)$-competitive
  algorithm. The same bounds apply for minimizing maximum weighted
  response time and maximum weighted delay factor.
\end{theorem}

\begin{remark}
  Minimizing maximum delay factor is NP-hard and there is no $(2-
  \eps)$-approximation unless $P=NP$ for any $\eps >0$ in the {\em
    offline} setting for unit-sized pages. There is a polynomial time
  computable $2$-speed schedule with the optimal delay factor (with
  $1$-speed) \cite{ChangEGK08}.  Theorem~\ref{thm:delayfactor} gives a
  polynomial time computable $(1+\eps)$-speed schedule that is
  $O(1/\eps^2)$-optimal (with $1$-speed).
\end{remark}

We remark that the algorithm $\sbg$ makes an interesting tradeoff between
two competing metrics and we explain this tradeoff in the context of
weighted response time and a lower bound we prove in this paper for
a simple greedy algorithm.
Recall that FIFO is $2$-competitive for maximum response time in
broadcast scheduling and is optimal for job scheduling. What are
natural ways to generalize FIFO to delay factor and weighted response
time? As shown in \cite{ChekuriM09}, $\sug$ (which is equivalent to
maximum weight first for weighted response time) is
$O(1/\eps)$-competitive with $(1+\eps)$-speed for job scheduling but
is not competitive for broadcast scheduling --- it may end up doing
much more work than necessary by transmitting a page repeatedly
instead of waiting and accumulating requests for a page. One natural
algorithm that extends FIFO for delay factor or weighted response time
is to schedule the request in the queue that has the largest current
delay factor (or weighted wait time). This greedy algorithm
was labeled $\lf$ (longest first) since it can be seen as an extension
of the well-studied $\lwf$ (longest-wait-first) for average flow
time. Since $\lwf$ is known to be $O(1)$-competitive with $O(1)$-speed
for average flow time, it was suggested in \cite{ChekuriIM09} that
$\lf$ may be $O(1)$-speed $O(1)$-competitive for maximum delay
factor. We show that this is not the case even for unicast scheduling.

\begin{theorem}
\label{thm:lf}
  For any constants $s, c > 1$, $\lf$ is not $c$-competitive with $s$-speed
  for minimizing maximum delay factor (or weighted response time)
  in unicast scheduling of unit-time jobs.
\end{theorem}

Our algorithm $\sbg$ can be viewed as an interesting tradeoff between
$\sug$ and $\lf$. $\sug$ gives preference to small slack requests
while the $\lf$ strategy helps avoid doing too much extra work in
broadcast scheduling by giving preference to pages that have waited
sufficiently long even if they have large slack. The algorithm $\sbg$
considers all requests whose delay factor at time $t$ (or weighted
wait time) is within a constant factor of the largest delay factor at
$t$ and amongst those requests schedules the one with the smallest
slack. This algorithmic principle may be of interest in other settings
and is worth exploring in the future.

\medskip
\noindent {\bf Other Related Work:} We have focussed on maximum response time and its variants and have already
discussed closely related work. Other metrics that have received substantial attention in broadcast scheduling are
minimizing average flow time and maximizing throughput of satisfied requests when requests have deadlines. We refer the
reader to a comprehensive survey on online scheduling algorithms by Pruhs, Sgall and Torng \cite{PruhsST}  (see also
\cite{Pruhs07}). The recent paper of Chang \etal \cite{ChangEGK08} addresses, among other things, the offline
complexity of several basic problems in broadcast scheduling. Average flow-time received substantial attention in both
the offline and online settings \cite{KalyanasundaramPV00,ErlebachH02,GandhiKKW04,GandhiKPS06,BansalCKN05,BansalCS06}.
For average flow time, there are three $O(1)$-speed $O(1)$-competitive online algorithms. $\lwf$ is one of them
\cite{EdmondsP03,ChekuriIM09} and the others are BEQUI \cite{EdmondsP03} and its extention \cite{EdmondsP09}. Our
recent work \cite{ChekuriIM09} has investigated $L_k$ norms of flow-time and showed that $\lwf$ is $O(k)$-speed
$O(k)$-competitive algorithm. Constant competitive online algorithms for maximizing throughput for unit-sized pages can
be found in \cite{Kimc04,ChanLTW04,ZhengFCCPW06,ChrobakDJKK06}. A more thorough description of related work is deferred
to a full version of the paper.

\medskip
\noindent {\bf Organization:} We prove each of the theorems mentioned
above in a different section. The algorithm and analysis for weighted
response time and weighted delay factor are very similar to that for
delay factor and hence, in this version, we omit the analysis and only
describe the algorithm.

\medskip
\noindent {\bf Notation:}
We denote the length of page $p$ by $\ell_p$. That is, $\ell_p$ is the
amount of time a 1-speed server takes to broadcast page $p$
non-preemptively.  We assume without loss of generality that for any
request $J_{p,i}$, $\st_{p,i} \ge \ell_p$. For an algorithm $A$ we let
$\alpha^A$ denote the maximum delay factor witnessed by $A$ for a
given sequence of requests.  We let $\alpha^*$ denote the optimal
delay factor of an offline schedule.  Likewise, we let $\rho^A$ denote
the maximum response time witnessed by $A$ and $\rho^*$ the optimal
response time of an offline schedule.  For a time interval $I=[a,b]$
we define $|I| = b - a$ to be the length of interval $I$.

\section{Minimizing the Maximum Response Time}

In this section we analyze $\fifo$ for minimizing maximum response
time when page sizes are different.  We first describe the algorithm
$\fifo$. $\fifo$ broadcasts pages {\em non-preemptively}.  Consider a
time $t$ when $\fifo$ finished broadcasting a page. Let $J_{p,i}$ be
the request in $\fifo$'s queue with earliest arrival time breaking
ties arbitrarily. $\fifo$ begins broadcasting page $p$ at time $t$.
At any time during this broadcast, we will say that $J_{p,i}$
\emph{forced} $\fifo$ to broadcast page $p$ at this time.  When
broadcasting a page $p$ all requests for page $p$ that arrived before
the start of the broadcast are simultaneously satisfied when the
broadcast completes.  Any request for page $p$ that arrive during the
broadcast are not satisfied until the next full transmission of $p$.

\newcommand{\cji}{\mathcal{J}_I}

We consider $\fifo$ when given a $1$-speed machine.  Let $\sigma$ be
an arbitrary sequence of requests.  Let $\opt$ denote some fixed
offline optimum schedule and let $\rho^*$ denote the optimum maximum
response time. We will show that $\rho^{\fifo} \leq 2 \rho^*$. For the
sake of contradiction, assume that $\fifo$ witnesses a response time
$c \rho^*$ by some job $J_{q,k}$ for some $c > 2$. Let $t^*$ be the
time $J_{q,k}$ is satisfied, that is $t^* = f_{q,k}$. Let $t_1$ be the
smallest time less than $t^*$ such that at any time $t$ during the
interval $[t_1, t^*]$ the request which forces $\fifo$ to broadcast a
page at time $t$ has response time at least $\rho^*$ when
satisfied. Throughout the rest of this section we let $I = [t_1,
t^*]$. Let $\cji$ denote the requests which forced $\fifo$ to
broadcast during $I$. Notice that during the interval $I$, all
requests in $\cji$ are completely satisfied during this interval. In
other words, any request in $\cji$ starts being satisfied during $I$
and is finished during $I$.

We say that $\opt$ \emph{merges} two distinct requests for a page $p$
if they are satisfied by the same broadcast.

\begin{lemma}
\label{lem:main-fifo} $\opt$ cannot merge any two requests in $\cji$ into a single broadcast.
\end{lemma}
\begin{proof}
  Let $J_{p,i}, J_{p,j} \in \cji$ such that $i < j$.  Note that that
  $J_{p,i}$ is satisfied before $J_{p,j}$. Let $t'$ be the time that
  $\fifo$ \emph{starts} satisfying request $J_{p,i}$. By the
  definition of $I$, request $J_{p,i}$ has response time at least
  $\rho^*$. The request $J_{p,j}$ must arrive after time $t'$, that is
  $a_{p,j} > t'$, otherwise request $J_{p,j}$ is satisfied by the
  same broadcast of page $p$ that satisfied $J_{p,i}$. Therefore, it
  follows that if $\opt$ merges $J_{p,i}$ and $J_{p,j}$ then the
  finish time of $J_{p,i}$ in $\opt$ is strictly greater than its
  finish time in $\fifo$ which is already at least $\rho^*$; this is
  a contradiction to the definition of $\rho^*$.
\end{proof}

\begin{lemma}
\label{lem:arrival-fifo} All requests in $\cji$ arrived no earlier than time $t_1 - \rho^*$.
\end{lemma}
\begin{proof}
  For the sake of contradiction, suppose some request $J_{p,i} \in
  \cji$ arrived at time $a_{p,i} < t_1 - \rho^*$.  During the interval
  $[a_{p,i} + \rho^*, t_1]$ the request $J_{p,i}$ must have wait time
  at least $\rho^*$.  However, then any request which forces $\fifo$
  to broadcast during $[a_{p,i} + \rho^*, t_1]$ must have response
  time at least $\rho^*$, contradicting the definition of $t_1$.
\end{proof}

We are now ready to prove Theorem~\ref{thm:fifo}, stating that $\fifo$
is 2-competitive.

\begin{proof}
  Recall that all requests in $\cji$ are completely satisfied during
  $I$. Thus we have that the total size of requests in $\cji$ is
  $|I|$. By definition $J_{q,k}$ witnesses a response time greater
  than $2\rho^*$ and therefore $t^* - a_{q,k} > 2\rho^*$. Since
  $J_{q,k} \in \cji$ is the last request done by $\fifo$ during $I$,
  all requests in $\cji$ must arrive no later than
  $a_{q,k}$. Therefore, these requests must be finished by time
  $a_{q,k} + \rho^*$ by the optimal solution. From
  Lemma~\ref{lem:arrival-fifo}, all the requests $\cji$ arrived no
  earlier than $t_1 - \rho^*$.  Thus $\opt$ must finish all requests
  in $\cji$, whose total volume is $|I|$, during $I_{opt} = [t_1 -
  \rho^*, a_{q,k}+\rho^*]$. Thus it follows that $|I| \leq |[t_1 -
  \rho^*, a_{q,k}+\rho^*]|$, which simplifies to $t^* \leq a_{q,k} +
  2\rho^*$. This is a contradiction to the fact that $t^* - a_{q,k} >
  2 \rho^*$.
\end{proof}

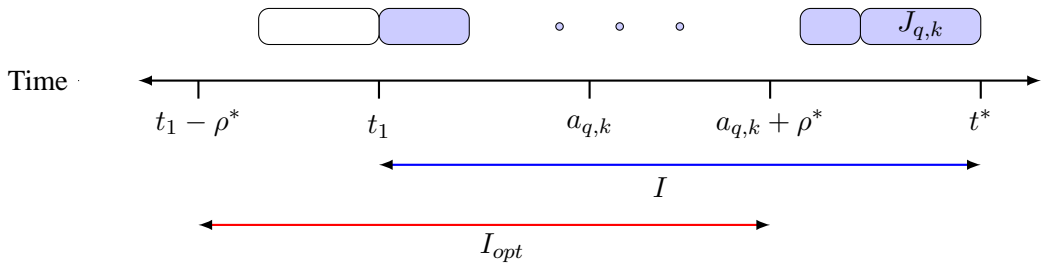
\begin{figure}[h]
    \figlab{fifo}

  \begin{center}

    \begin{tikzpicture}[scale = 0.8]

        \tikzstyle{opt} = [>=latex,draw=red,fill=red, thick];
        \tikzstyle{lf} = [>=latex,draw=blue,fill=blue, thick];
        \tikzstyle{th} = [>=latex,draw=black, thick];
        \tikzstyle{rec1} = [>=latex,draw=black,rounded corners, fill = blue!20, thin];
        \tikzstyle{rec2} = [>=latex,draw=black, rounded corners,fill = white!20, thin];

      \tikzstyle{vtx}=[circle,draw, fill = blue!20, inner sep=1pt];

      \begin{scope}
        \path (0, -0.3) coordinate (vn);  
        \path (0, -1.4) coordinate (vi1);  
        \path (0, -2.4) coordinate (vi2);  

        \path (-14,0) coordinate (x0);
        \path (-13,0) coordinate (t1_rhos);   
        \path (-10,0) coordinate (t1);       
        \path (-6.5,0) coordinate (a);       
        \path (-3.5,0) coordinate (aprho);    
        \path (0,0) coordinate (ts);     
        \path (1,0) coordinate (x1);

       \draw[th, <->] (x0) -- (x1);
        \draw[th] (t1_rhos) -- +(vn)    node[below] {$t_1 - \rho^*$};
        \draw[th] (t1) -- +(vn)         node[below=0.08cm] {$t_1$};
        \draw[th] (a) -- +(vn)         node[below=0.08cm] {$a_{q,k}$};
        \draw[th] (aprho) -- +(vn)         node[below] {$a_{q,k} + \rho^*$};
        \draw[th] (ts) -- +(vn)         node[below] {$t^*$};

       \draw (-15,0) -- (-15,0) node[left] {Time};

        \path (t1) -- +(vi1) coordinate (Ib);    \path (ts) -- +(vi1) coordinate (If);
       \draw[<->,lf] (Ib) -- (If) node[below = 0.3cm , right=-4.5cm] {$I$};
        \path (t1_rhos) -- +(vi2) coordinate (Ob);    \path (aprho) -- +(vi2) coordinate (Of);
       \draw[<->,opt] (Ob) -- (Of) node[below = 0.3cm, right=-4cm] {$I_{opt}$};

        \path (0, 1) coordinate (h);  

        \draw[rec1] (-2,0.6) rectangle (0,1.2) node[right = -0.8cm, below = -0.1cm] {$J_{q,k}$};

        \draw[rec1] (-3,0.6) rectangle (-2,1.2);
        \node (s0) at (-7,0.9) [vtx] {};
        \node (s1) at (-6,0.9) [vtx] {};
        \node (s2) at (-5,0.9) [vtx] {};

        \draw[rec1] (-10, 0.6) rectangle (-8.5,1.2);
        \draw[rec2] (-12, 0.6) rectangle (-10,1.2);

      \end{scope}

    \end{tikzpicture}

  \end{center}

  \vspace{-0.75cm}
  \caption{ Broadcasts by $\fifo$ satisfying requests in $\cji$ are shown in blue. Note that $a_{q,k}$ and $a_{q,k}+ \rho^*$ are not necessarily contained in $I$.}
\end{figure}

We now discuss the differences between our proof of $\fifo$ for varying sized pages and the proof given by Chang \etal
in \cite{ChangEGK08} showing that $\fifo$ is 2-competitive for unit sized pages.  In \cite{ChangEGK08} it is shown that
at anytime $t$, $F(t)$, the set of {\em unique} pages in $\fifo$'s queue satisfies the following property: $|F(t)
\setminus O(t)| \le |O(t)|$ where $O(t)$ is the set of unique pages in $\opt$'s queue. This easily implies the desired
bound. To establish this, they use a slot model in which unit-sized pages arrive only during integer times which allows
one to define unique pages. This may appear to be a technicality, however when considering different sized pages, it is
not so clear how one even defines unique pages since this number varies during the transmission of $p$ as requests
accumulate. Our approach
avoids this issue in a clean manner by not assuming a slot model or unit-sized pages.

\section{Minimizing Maximum Delay Factor and Weighted Response time}

In this section we consider the problem of minimizing maximum delay
factor and prove Theorem~\ref{thm:delayfactor}.

\subsection{Unit Sized Pages}
In this section we consider the problem of minimizing the maximum
delay factor when all pages are of unit size. In this setting we
assume preemption is not allowed.  In the standard unicast scheduling
setting where each broadcast satisfies exactly one request, it is
known that the algorithm which always schedules the request with
smallest slack at any time is $(1+\eps)$-speed
$O(\frac{1}{\eps})$-competitive \cite{ChekuriM09}.  However, in the
broadcast setting this algorithm, along with other simple greedy
algorithms, do not provide constant competitive ratios even with extra
speed.  The reason for this is that the adversary can force these
algorithm to repeatedly broadcast the same page even though the
adversary can satisfy each of these requests in a singe broadcast.

Due to this, we consider a more sophisticated algorithm called $\sbg$ (Shortest-Slack-First with Waiting). This
algorithm was developed and analyzed in \cite{ChekuriM09}. In this paper we alter the algorithm in a slight but
practically important way. The main contribution is, however, a new analysis that is at a high-level similar in outline
to the one in \cite{ChekuriM09} but is subtly different and leads to much improved bound on its performance.  $\sbg$
\emph{adaptively} forces requests to wait after their arrival before they are considered for scheduling.  The algorithm
is parameterized by a real value $c > 1$ which is used to determine how long a request should wait.  Before scheduling
a page at time $t$, the algorithm determines the largest current delay factor of any request that is unsatisfied at
time $t$, $\alpha_t$.  Amongst the unsatisfied requests that have a current delay factor at least $\frac{1}{c}
\alpha_t$, the page corresponding to the request with smallest slack is broadcasted. Note that in the algorithm, each
request is forced to wait to be scheduled until it has delay factor at least $\frac{1}{c} \alpha_t$.  Thus $\sbg$ can
be seen an adaptation of the algorithm which schedules the request with smallest slack in broadcasting setting with
explicit waiting.  Waiting is used to potentially satisfy multiple requests with similar arrival times in a single
broadcast.  Another interpretation, that we mentioned earlier, is that $\sbg$ is a balance between $\lf$ and $\sug$.

\begin{center}
\begin{tabular}[r]{|c|}
\hline
\textbf{Algorithm}: \sbg \\
\\

\begin{minipage}{\textwidth}
\begin{itemize}
\item Let $\alpha_t$ be the maximum delay factor of any request in
 $\sbg$'s queue  at time $t$.

\item At time $t$,  let $Q(t) = \{ J_{p,i} \mid \mbox{ $J_{p,i}$
has not been satisfied and ${{t - a_{p,i} \over {\st_{p,i}}} } \geq \frac{1}{c} \alpha_t $} \}$.

\item If the machine is free at $t$, schedule the request in
$Q(t)$ with the smallest slack {\em non-preemptively}.
\end{itemize}
\end{minipage}\\

\hline
\end{tabular}
\end{center}

First we note the difference between $\sbg$ above and the one described in \cite{ChekuriM09}.  Let $\alpha'_t$ denote
the maximum delay factor witnessed so far by $\sbg$ at time $t$ over all requests seen by $t$ including satisfied and
unsatisfied requests. In \cite{ChekuriM09}, a request $J_{p,i}$ is in $Q(t)$ if $\frac{t-a_{p,i}}{\st_{p,i}} \geq
\frac{1}{c}\alpha'_t$. Note that $\alpha'_t$ is monotonically increases with $t$ while $\alpha_t$ can increase and
decrease with $t$ and is never more than $\alpha'_t$. In the old algorithm it is possible that $Q(t)$ is empty and no
request is scheduled at $t$ even though there are outstanding requests! Our new version of $\sbg$ can be seen as more
practical since there will always be requests in $Q(t)$ if there are outstanding requests and moreover it adapts and
may reduce $\alpha_t$ as the request sequence changes with time.  It is important to note that our analysis and the
analysis given in $\cite{ChekuriM09}$ hold for both definitions of $\sbg$ with some adjustments.

We analyze $\sbg$ when it is given a $(1 + \eps)$-speed machine. Let $c > 1 + \frac{2}{\eps}$ be the constant which
parameterizes $\sbg$. Let $\sigma$ be an arbitrary sequence of requests. We let $\opt$ denote some fixed offline
optimum schedule and let $\alpha^*$ and $\alpha^{\sbg}$ denote the maximum delay factor achieved by $\opt$ and $\sbg$,
respectively. We will show that $\alpha^{\sbg} \le c^2\alpha^*$.  For the sake of contradiction, suppose that $\sbg$
witnesses a delay factor greater than $c^2 \alpha^*$. We consider the \emph{first} time $t^*$ when $\sbg$ has some
request in its queue with delay factor $c^2 \alpha^{*}$. Let the request $J_{q,k}$ be a request which achieves the
delay factor $c^2 \alpha^{*}$ at time $t^*$. Let $t_1$ be the smallest time less than $t^*$  such that at each time $t$
during the interval $[t_1, t^*]$ if $\sbg$ is forced to broadcast by request $J_{p,i}$ at time $t$ it is the case that
$\frac{t - a_{p,i}}{S_{p,i}} \geq \alpha^*$ and $\st_{p,i} \leq \st_{q,k}$. Throughout this section we let $I = [t_1, t^*]$. The main difference between the analysis in
\cite{ChekuriM09} and the one here is in the definition of $t_1$. In \cite{ChekuriM09}, $t_1$ was implicitly defined to
be $a_{q,k} + c(f_{q,k} - a_{q,k})$.

We let $\cji$ denote the requests which forced $\sbg$ to schedule
broadcasts during the interval $[t_1, t^*]$. We now show that any two
request in $\cji$ cannot be satisfied with a single broadcast by the
optimal solution. Intuitively, the most effective way the adversary to
performs better than $\sbg$ is to merge requests of the same page into
a single broadcast.  Here we will show this is not possible for the
requests in $\cji$. We defer the proof of Lemma~\ref{lem:main} to the
Appendix, since the proof is similar to that of
Lemma~\ref{lem:main-fifo}

\begin{lemma}
  \label{lem:main}
    $\opt$ cannot merge any two requests in $\cji$ into a single broadcast.
\end{lemma}

To fully exploit the advantage of speed augmentation, we need to
ensure that the length of the interval $I$ is sufficiently long.

\begin{lemma}
    \label{lem:len}
    $|I| = |[t_1, t^*]| \geq (c^2 -c)\st_{q,k}\alpha^*$.
\end{lemma}
\begin{proof}
  The request $J_{q,k}$ has delay factor at least $c\alpha^*$ at any time during $I' = [t', t^*]$, where
  $t' = t^* -(c^2 -c)\st_{q,k}\alpha^*$. Let $\tau \in I'$.
  The largest delay factor any request can have at time $\tau$ is less than $c^2 \alpha^*$ by definition of
  $t^*$ being the first time $\sbg$ witnesses delay factor $c^2
  \alpha^*$. Hence, $\alpha_{\tau} \leq c^2 \alpha^*$.  Thus, the
  request $J_{q,k}$ is in the queue $Q(\tau)$ because $c\alpha^* \geq
  \frac{1}{c}\alpha_{\tau}$.  Moreover, this means that any request that
  forced $\sbg$ to broadcast during $I'$, must have delay
  factor at least $\alpha^*$ and since $J_{q,k} \in Q(\tau)$ for any $\tau \in I'$, the
  requests scheduled during $I'$ must have slack at most
  $\st_{q,k}$.
\end{proof}

We now explain a high level view of how we lead to a
contradiction. From Lemma~\ref{lem:main}, we know any two requests in
$\cji$ cannot be merged by $\opt$. Thus if we show that $\opt$ must
finish all these requests during an interval which is not long enough
to include all of them, we can draw a contradiction. More precisely,
we will show that all requests in $\cji$ must be finished during
$I_{opt}$ by $\opt$, where $I_{opt} = [t_1 -2\st_{q,k} \alpha^* c,
t^*]$. It is easy to see that all these requests already have delay
factor $\alpha^*$ by time $t^*$, thus the optimal solution must finish
them by time $t^*$. For the starting point, we will bound the arrival
times of the requests in $\cji$ in the following lemma. After that, we
will draw a contradiction in \lemref{competitiveone}.

\begin{lemma}
  \label{lem:arrivaltime}
  Any request in $\cji$ must have arrived after time $t_1 - 2
  \st_{q,k} \alpha^* c$.
\end{lemma}

\begin{proof}
  For the sake of contradiction, suppose that some request $J_{p,i}
  \in \cji$ arrived at time $t' < t_1 -2 \st_{q,k} \alpha^* c$. Recall
  that $J_{p,i}$ has a slack no bigger than $S_{q,k}$ by the
  definition of $I$. Therefore at time $t_1 - \st_{q,k} \alpha^* c$,
  $J_{p,i}$ has a delay factor of at least $c \alpha^*$. Thus any
  request scheduled during the interval $I' = [t_1 - \st_{q,k}
  \alpha^* c, t_1]$ has a delay factor no less than $\alpha^*$. We
  observe that $J_{p,i}$ is in $Q(\tau)$ for $\tau \in I'$; otherwise
  there must be a request with a delay factor bigger than $c^2
  \alpha^*$ at time $\tau$ and this is a contradiction to the
  assumption that $t^*$ is the first time that $\sbg$ witnessed a
  delay factor of $c^2 \alpha^*$. Therefore any request scheduled
  during $I'$ has a slack no bigger than $S_{p,i}$. Also we know that
  $S_{p,i} \leq S_{q,k}$. In sum, we showed that any request done
  during $I'$ had slack no bigger than $S_{q,k}$ and a delay factor no
  smaller than $\alpha^*$, which is a contradiction to the definition
  of $t_1$.
\end{proof}

Now we are ready to prove the competitiveness of $\sbg$.

\begin{lemma}
  \lemlab{competitiveone}
Suppose $c$ is a constant s.t. $ c >  1 + 2/ \eps$. If $\sbg$ has $(1+ \eps)$-speed then
  $\alpha^{\sbg} \leq c^2 \alpha^*$.
\end{lemma}

\begin{proof}
For the sake of contradiction, suppose that $\alpha^{\sbg} > c^2 \alpha^*$. During the interval $I $, the number of
broadcasts which $\sbg$ transmits is $(1+\eps)|I|$. From Lemma~\ref{lem:arrivaltime}, all the requests processed during
$I$ have arrived no earlier than $t_1 -2 c \alpha^* \st_{q,k}$. We know that the optimal solution must process these
requests before time $t^*$ because these requests have delay factor at least $\alpha^*$ by $t^*$. By
Lemma~\ref{lem:main} the optimal solution must make a unique broadcast for each of these requests. Thus, the optimal
solution must finish all of these requests in $2 c \alpha^* \st_{q,k} + |I|$ time steps. Thus, the it must hold that
$(1+\eps)|I| \leq 2c \alpha^* \st_{q,k} + |I|$. Using Lemma~\ref{lem:len}, this simplifies to $c \leq 1 + 2/\eps$,
which is a contradiction to $c > 1 + 2/\eps$,.
\end{proof}

The previous lemmas prove the first part of
Theorem~\ref{thm:delayfactor} when $c = 1 + 3/\eps$. Namely that
$\sbg$ is a $(1+ \epsilon)$-speed
$O(\frac{1}{\epsilon^2})$-competitive algorithm for minimizing the
maximum delay factor in broadcast scheduling with unit sized pages.

We now compare proof of Theorem~\ref{thm:delayfactor} and the proof of Theorem~\ref{thm:fifo} with the analysis given
in $\cite{ChekuriM09}$. The central technique used in $\cite{ChekuriM09}$ and in our analysis is to draw a
contradiction by showing that the optimal solution must complete more requests than possible on some time interval $I$.
This technique is well known in unicast scheduling.  At the heart of this technique is to find the which requests to
consider and bounding the length of the interval $I$.  This is where our proof and the one given in \cite{ChekuriM09}
differ. Here we are more careful on how $I$ is defined and how we find requests the optimal solution must broadcast
during $I$. This allows us to show tighter bounds on the speed and competitive ratios while simplifying the analysis.
In fact, our analysis of $\fifo$ and $\sbg$ shows the importance of these definitions. Our analysis of $\fifo$ shows
that a tight bound on the length of $I$ can force a contradiction without allowing extra speed-up given to the
algorithm.  Our analysis of $\sbg$ shows that when the length of $I$ varies how resource augmentation can be used to
force the contradiction.

\subsection{Weighted Response Time and Weighed Delay Factor}
Before showing that $\sbg$ is $(2+\epsilon)$-speed
$O(\frac{1}{\epsilon^2})$-competitive for minimizing the maximum delay
factor with different sized pages, we show the connection of our
analysis of $\sbg$ to the problem of minimizing \emph{weighted}
response time.  In this setting a request $J_{p,i}$ has a weight
$w_{p,i}$ instead of a slack.  The goal is to minimize the maximum
weighted response time $\max_{p,i} w_{p,i}(f_{p,i} - a_{p,i})$.  We
develop an algorithm which we call $\bwf$ for Biggest-Wait-First with
Waiting.  This algorithm is defined analogously to the definition of
$\sbg$. The algorithm is parameterized by a constant $c >1$.  At any
time $t$ before broadcasting a page, $\bwf$ determines the largest
weighted wait time of any request which has yet to be satisfied. Let
this value be $\rho_t$.  The algorithm then chooses to broadcast a
page corresponding to the request with largest weight amongst the
requests whose current weighted wait time at time $t$ is larger than
$\frac{1}{c}\rho_t$.

\begin{center}
\begin{tabular}[r]{|c|}
\hline
\textbf{Algorithm}: \bwf \\
\\

\begin{minipage}{\textwidth}
\begin{itemize}
\item Let $\rho_t$ be the maximum weighted wait time of any request in
 $\bwf$'s queue  at time $t$.

\item At time $t$,  let $Q(t) = \{ J_{p,i} \mid \mbox{ $J_{p,i}$
has not been satisfied and $w_{p,i}(t - a_{p,i}) \geq \frac{1}{c} \rho_t $} \}$.

\item If the machine is free at $t$, schedule the request in
$Q(t)$ with largest weight {\em non-preemptively}.
\end{itemize}
\end{minipage}\\

\hline
\end{tabular}
\end{center}

Although minimizing the maximum delay factor and minimizing the
maximum weighted flow time are very similar metrics, the problems are
not equivalent.

It may also be of interest to minimize the maximum \emph{weighted}
delay factor.  In this setting each request has a deadline and a
weight.  The goal is to minimize $\max_{p,i} w_{p,i}(f_{p,i} -
a_{p,i}) / S_{p,i}$.  For this setting we develop another algorithm
which we call $\srfw$ (Smallest-Ratio-First with Waiting).  The
algorithm takes the parameter $c$. At any time $t$ before broadcasting
a page, $\srfw$ determines the largest weighted delay factor of any
request which has yet to be satisfied. Let this value be $\alpha^w_t$.
The algorithm then chooses to broadcast a page corresponding to the
request with the smallest ratio of the slack over the weight amongst
the requests whose current weighted delay factor at time $t$ is larger
than $\frac{1}{c}\alpha^w_t$.  The algorithm can be formally expressed
as follows.

\begin{center}
\begin{tabular}[r]{|c|}
\hline
\textbf{Algorithm}: \srfw \\
\\

\begin{minipage}{\textwidth}
\begin{itemize}
\item Let $\alpha^w_t$ be the maximum weighted delay factor of any request in
 $\srfw$'s queue  at time $t$.

\item At time $t$, let $Q(t) = \{ J_{p,i} \mid \mbox{ $J_{p,i}$ has
    not been satisfied and $w_{p,i}(t - a_{p,i})/\st_{p,i} \geq
    \frac{1}{c} \alpha^w_t $} \}$.

\item If the machine is free at $t$, schedule the request in
$Q(t)$ with smallest slack over weight {\em non-preemptively}.
\end{itemize}
\end{minipage}\\

\hline
\end{tabular}
\end{center}

For the problems of minimizing the maximum weighted response time and
weighted delay factor, the upper bounds shown for $\sbg$ in this paper
also hold for $\bwf$ and $\srfw$, respectively.  The analysis of
$\bwf$ and $\srfw$ is very similar to that of $\sbg$ and the proofs
are omitted.

\subsection{Varying Sized Pages}

Here we extend our ideas to the case where pages can have a different
sizes for the objective of minimizing the maximum delay factor.  We
develop a generalization of $\sbg$ for this setting which is similar
to the generalization of $\sbg$ given in \cite{ChekuriM09}.  For each
page $p$, we let $\ell_p$ denote the length of page $p$.  Since pages
have different lengths, we allow preemption. Therefore, if $t_1$ is
the time where the broadcast of page $p$ is started and $t_2$ is the
time that this broadcast is completed it is the case that $\ell_p \leq
t_2 - t_1$.  A request for the page $p$ is satisfied by this broadcast
only if the request arrives before time $t_1$.  A request that arrives
during the interval $(t_1, t_2]$ does not start being satisfied
because it must receive a sequential transmission of page $p$ starting
from the beginning.  It is possible that a transmission of page $p$ is
$\emph{restarted}$ due to another request for page $p$ arriving which
has smaller slack. The original transmission of page $p$ in this case
is abandoned. Notice that this results in wasted work by the
algorithm.  It is because of this wasted work that more speed is
needed to show the competitiveness of $\sbg$.

We outline the details of modifications to $\sbg$. As before, at any
time $t$, the algorithm maintains a queue $Q(t)$ at each time where a
request $J_{p,i}$ is in $Q(t)$ if and only if $\frac{t -
  a_{p,i}}{\st_{p,i}} \geq \frac{1}{c} \alpha_t$. The algorithm
broadcasts a request with the smallest slack in $Q(t)$. The algorithm
may preempt a broadcast of $p$ that is forced by request $J_{p,i}$ if
another request $J_{p',j}$ becomes available for scheduling such that
$S_{p',j} < S_{p,i}$.  If the request $J_{p,i}$ ever forces $\sbg$ to
broadcast again, then $\sbg$ continues to broadcast page $p$ from
where it left off before the preemption.  If another request for page
$p$ forces $\sbg$ to broadcast page $p$ before $J_{p,i}$ is satisfied,
then the transmission of page $p$ is $\emph{restarted}$. A key
difference between our generalization of $\sbg$ and the one from
\cite{ChekuriM09} is that in our new algorithm, requests can be forced
out of $Q$ even after they have been started. Hence, in our version of
$\sbg$ every request in $Q(t)$ has current delay factor at least
$\frac{1}{c} \alpha_t$ at time $t$. Our algorithm breaks ties
arbitrarily. In \cite{ChekuriM09}, ties are broken arbitrarily, but
the algorithm ensures that if a request $J_{p,k}$ is started before a
request $J_{p',j}$ then $J_{p,k}$ will be finished before request
$J_{p',j}$. Here, this requirement is not needed. Note that the
algorithm may preempt a request $J_{p,i}$ by another request
$J_{p,k}$, for the same page $p$ if $\st_{p,k} < \st_{p,i}$. In this
case the first broadcast of page $p$ is abandoned. Notice that
multiple broadcasts of page $p$ can repeatedly be abandoned.

We now analyze the extended algorithm assuming that it has a
$(2+\eps)$-speed advantage over the optimal offline algorithm. As
mentioned before, the extra speed is needed to overcome the wasted
work by abandoning broadcasts.

As before, let $\sigma$ be an arbitrary sequence of requests.  We let $\opt$ denote some fixed offline optimum schedule
and let $\alpha^*$ denote the optimum delay factor. Let $c > 1 + \frac{4}{\eps}$ be the constant that parameterizes
$\sbg$.  We will show that $\alpha^{\sbg} \le c^2\alpha^*$.  For the sake of contradiction, suppose that $\sbg$
witnesses a delay factor greater than $c^2 \alpha^*$. We consider the \emph{first} time $t^*$ when $\sbg$ has some
request in its queue with delay factor $c^2 \alpha^{*}$. Let the request $J_{q,k}$ be a request which achieves the
delay factor $c^2 \alpha^{*}$ at time $t^*$. Let $t_1$ be the smallest time less than $t^*$  such that at each time $t$
during the interval $[t_1, t^*]$ if $\sbg$ is forced to broadcast by request $J_{p,i}$ at time $t$ it is the case that
$\frac{t - a_{p,i}}{S_{p,i}} \geq \alpha^*$ and $\st_{p,i} \leq \st_{q,k}$. Throughout this section we let $I=[t_1,
t^*]$. Notice that some requests that force $\sbg$ to broadcast during $I$ could have started being satisfied before
$t_1$.

We say that a request \emph{starts} being scheduled at time $t$ if it
is the request which forces $\sbg$ to broadcast at time $t$ and $t$ is
the first time the request forces $\sbg$ to schedule a page. Notice
that a request can only start being satisfied once and at most one
request starts being scheduled at any time.  We now show a lemma
analogous to Lemma~\ref{lem:main}.

\begin{lemma}
  \label{lem:main-varying} Consider two distinct requests $J_{x,j}$
  and $J_{x,i}$ for some page $x$. If $J_{x,j}$ and $J_{x,i}$ both
  start being scheduled by $\sbg$ during the interval $I$ then $\opt$
  cannot satisfy $J_{x,j}$ and $J_{x,i}$ by a single broadcast.
\end{lemma}
\begin{proof}
  Without loss of generality say that request $J_{x,j}$ was satisfied
  before request $J_{x,i}$ by $\sbg$.  Let $t'$ be the time that
  $\sbg$ \emph{starts} satisfying request $J_{x,j}$.  By the
  definition of $I$, request $J_{x,j}$ must have delay factor at least
  $\alpha^*$ at this time.  We also know that the request $J_{x,i}$
  must arrive after time $t'$, otherwise request $J_{x,i}$ must also
  be satisfied at time $t'$.  If the optimal solution combines these
  requests into a single broadcast then the request $J_{x,j}$ must
  wait until the request $J_{x,i}$ arrives to be satisfied. However,
  this means that the request $J_{x,j}$ must achieve a delay factor
  greater than $\alpha^*$ by $\opt$, a contradiction of the definition
  of $\alpha^*$.
\end{proof}

The next two lemmas have proofs similar to Lemma~\ref{lem:len} and
Lemma~\ref{lem:arrivaltime}, we defer the proofs to the Appendix.

\begin{lemma}
    \label{lem:len-varying}
    $|I| = |[t_1, t^*]| \geq (c^2 -c)\st_{q,k}\alpha^*$.
\end{lemma}

\begin{lemma}
  \label{lem:arrivaltime-varying}
  Any request which forced $\sbg$ to schedule a page during $I$ must
  have arrived after time $t_1 - 2 \st_{q,k} \alpha^* c$.
\end{lemma}

Using the previous lemmas we can bound the competitiveness of $\sbg$. In the following lemma the main difference
between the proof for unit sized pages and the proof for varying sized pages can be seen.  The issue is that there can
be some requests which start being satisfied before time $t_1$ which force $\sbg$ to broadcast a page during the
interval $I$.  When these requests were started, their delay factor need not be bounded by $\alpha^*$.  Due to this, it
is possible for these requests to be merged with other requests which forced $\sbg$ to broadcast on the interval $I$.

\begin{lemma}
  \lemlab{competitive} Suppose $c$ is a constant s.t. $ c > 1 + 4/
  \eps$. If $\sbg$ has $(2+ \eps)$-speed then $\alpha^{\sbg} \leq c^2
  \alpha^*$.
\end{lemma}

\begin{proof}
  For the sake of contradiction, suppose that $\alpha^{\sbg} > c^2
  \alpha^*$. Let $\partially$ be the set of requests which start being
  satisfied before time $t_1$ which force $\sbg$ to broadcast at some
  time during $I$.  Notice that no two requests in $\partially$ are
  for the same page. Let $\zeroly$ be the set of requests which start
  being satisfied during the interval $I$.  Note that the sets
  $\partially$ and $\zeroly$ may consist of requests whose
  corresponding broadcast was abandoned at some point and that
  $\partially \cap \zeroly = \emptyset$ by definition. Let
  $V_\partially$ and $V_\zeroly$ denote the total sum of size of the
  requests in $\partially$ and $\zeroly$, respectively.

  During the interval $I$, the volume of broadcasts which $\sbg$
  transmits is $(2+\eps)|I|$. Notice that $V_\partially + V_\zeroly
  \geq (2+\eps)|I|$, since $\partially \cup \zeroly$ accounts for all
  requests which forced $\sbg$ to broadcast their pages during
  $I$. From Lemma~\ref{lem:arrivaltime-varying}, \emph{all} the
  requests processed during $I$ have arrived no earlier than $t_1 -2 c
  \alpha^* \st_{q,k}$. We know that the optimal solution must process
  these requests before time $t^*$ because these requests have delay
  factor at least $\alpha^*$ by this time.

  By Lemma~\ref{lem:main-varying} the optimal solution must make a
  unique broadcast for each request in $\zeroly$.  We also know that
  no two requests in $\partially$ can be merged because no two
  requests in $\partially$ are for the same page. The optimal
  solution, however, could possibly merge requests in $\partially$
  with requests in $\zeroly$.  Thus, the optimal solution must
  broadcast at least a $\max\{V_\partially, V_\zeroly\}$ volume of requests
  during the interval $[t_1 - 2c\alpha^* \st_{q,k}, t^*]$.  Notice
  that $\max\{V_\partially, V_\zeroly \} \geq \frac{1}{2}
  (V_\partially + V_\zeroly) \geq \frac{1}{2} (2+\epsilon)|I|$ and
  that $|[t_1 - 2c\alpha^* \st_{q,k}, t^*]| = 2 c \alpha^* \st_{q,k} +
  |I|$. Therefore, it must hold that $\frac{1}{2}(2+\eps)|I| \leq 2c
  \alpha^* \st_{q,k} + |I|$. With Lemma~\ref{lem:len-varying}, this
  simplifies to $c \leq 1 + 4 / \eps$. This is a contradiction to $c >
  1 + 4/ \eps$.
\end{proof}

Thus, we have the second part of Theroem~\ref{thm:delayfactor} by
setting $c = 1 + 5/\eps$.  Namely that $\sbg$ is $(2+\epsilon)$-speed
$O(\frac{1}{\epsilon^2})$-competitive for minimizing the maximum delay
factor for different sized pages.

\section{Lower Bound for a Natural Greedy Algorithm $\lf$}

\newcommand{\cj}{\mathcal{J}}
\newcommand{\cs}{\mathcal{S}}

In this section, we consider a natural algorithm which is similar to $\sbg$.  This algorithm, which we will call $\lf$
for Longest Delay First, always schedules the page which has the largest delay factor. Notice that $\lf$ is the same as
$\sbg$ when $c=1$.  However, we are able to show a negative result on the algorithm for minimizing the maximum delay
factor.  This demonstrates the importance of the tradeoff between scheduling a request with smallest slack and forcing
requests to wait. The algorithm $\lf$ was suggested and analyzed in our recent work \cite{ChekuriIM09} and is inspired
by $\lwf$ which was shown to be $O(1)$-competitive with $O(1)$-speed for average flow time \cite{EdmondsP04}. In
\cite{ChekuriIM09} $\lf$ is shown to be $O(k)$-competitive with $O(k)$-speed for $L_k$ norms of flow time and delay
factor in broadcast scheduling for unit sized pages. Note that $\lf$ is a simple greedy algorithm. It was suggested in
\cite{ChekuriIM09} that $\lf$ may be competitive for maximum delay factor which is the $L_{\infty}$-norm of delay
factor.

To show the lower bound on the $\lf$, we will show that it is not $O(1)$-speed $O(1)$-competitive, even in the standard
unicast scheduling setting with unit sized jobs.  Since we are considering the unicast setting where processing a page
satisfies exactly one request, we drop the terminology of `requests' and use `jobs'.  We also drop the index of a
request $J_{p,i}$ and use $J_i$ since there can only be one request for each page.  Let us say that $J_{i}$ has a wait
ratio of $r_{i}(t) = \frac{t- a_{i}}{S_{i}}$ at time $t > a_i$, where $a_i$ and $S_i$ is the arrival time and slack
size of $J_i$. Note that the delay factor of $J_i$ is $\max(1, r_i(f_i))$ where $f_i$ is $J_i$'s finish time. We now
formally define $\lf$. The algorithm $\lf$ schedules the request with the largest wait ratio at each time.  $\lf$ can
be seen as a natural generalization of $\fifo$.  This is because $\fifo$ schedules the request with largest wait time
at each time. Recall that $\sbg$ forces requests to wait to help merge potential requests in a single broadcast.  The
algorithm $\lf$ behaves similarly since it implicitly delays each request until it is the request with the largest wait
ratio, potentially merging many requests into a single broadcast.  Hence, this algorithm is a natural candidate for the
problem of minimizing the maximum delay factor and it does not need any parameters like the algorithm $\sbg$. However,
this algorithm cannot have a constant competitive ratio with any constant speed.

For any speed-up $s \geq 1$ and any constant $c \geq 2$, we construct the following adversarial instance $\sigma$. For
this problem instance we will show that $\lf$ has wait ratio at least $c$, while $\opt$ has wait ratio at most $1$.
Hence, we can force $\lf$ to have a competitive ratio of $c$, for any constant $c \geq 2$. In the instance $\sigma$,
there are a series job groups $\cj_i$ for $0 \leq i \leq k$, where $k$ is a constant to be fixed later.  We now fix the
jobs in each group.  For simplicity of notation and readability, we will allow jobs to arrive at negative times. We can
simply shift each of the times later, so that all arrival times are positive. It is also assumed that $s$ and $c$ are
integers in our example.

All jobs in each group $\cj_i$ have the same arrival time $A_i =
-(sc)^{k-i+1} - \sum_{j=0}^{k-i-1} (sc)^j$ and have the same slack
size of $S_i = \frac{c(sc)^{k-i} }{ (1-1/sc)^{k-i}}$. There will be
$s(sc)^{k+1}$ jobs in the group $\cj_0$ and $s(sc)^{k-i}$ jobs in the
group $\cj_i$ for $1 \leq i \leq k$.

We now explain how $\lf$ and $\opt$ behave for the instance $\sigma$. For simplicity, we will refer to $\cj_i$, instead
of a job in $\cj_i$, since all jobs in the same group are indistinguishable to the schedular. For the first group
$\cj_0$, $\lf$ starts and keeps processing $\cj_0$ upon its arrival until completing it. On the other hand, we let
$\opt$ procrastinate $\cj_0$ until $\opt$ finishes all jobs in $\cj_1$ to $\cj_k$. This does not hurt $\opt$, since the
slack size of the jobs in $\cj_0$ is so large.  In fact, we will show that $\opt$ can finish $\cj_0$ by its deadline.
For each group $\cj_i$ for $1 \leq i \leq k$, $\opt$ will start $\cj_i$ upon its arrival and complete each job in
$\cj_i$ without interruption. To the contrary, for each $1 \leq i \leq k$ $\lf$ will not begin scheduling $\cj_i$ until
the jobs have been substantially delayed. The delay of $\cj_k$ is critical for $\lf$, since the slack of $\cj_k$ is
small. For intuitive understanding, we refer the reader to Figure 2. 

\begin{figure}[h]
    \figlab{groups}

  \begin{center}

{\footnotesize
    \begin{tikzpicture}[scale = 0.4]
         \tikzstyle{opt}=[>=latex,draw=red,fill=red, thick];
         \tikzstyle{lf}=[>=latex,draw=blue,fill=blue, thick];
{\footnotesize}
      \tikzstyle{vtx}=[circle,draw,inner sep=0.5pt];

      \begin{scope}

        \path (0,-0.3) coordinate (vn); 
        \path (0,0.2) coordinate (vg); 
        \path (0,2.5) coordinate (vg1); 
        \path (0,5) coordinate (vg2); 

        \path (-4,0) coordinate (A0); \path (-2,0) coordinate (O0); 
        \path (-1,0) coordinate (L0); \path (0,0) coordinate (F0);

        \path (-17,0)++(vg1) coordinate (A1); \path (-9,0)++(vg1) coordinate (O1); 
        \path (-5,0)++(vg1) coordinate (L1); \path (-1,0)++(vg1) coordinate (F1);

        \path (-23,0)++(vg2) coordinate (A2); \path (-23,0)++(vg2) coordinate (O2);   
        \path (-21,0)++(vg2) coordinate (L2); \path (-5,0)++(vg2) coordinate (F2);   

       \draw (A0) -- (O0) -- (L0) -- (F0);
       \draw (A0) -- +(vn) node[below] {\footnotesize{$A_k$}};
       \draw (O0) -- +(vn); 
       \draw (L0) -- +(vn); 
       \draw (F0) -- +(vn) node[below] {$F_k$};;
       \path (A0)++(vg) coordinate (aA0);
       \path (O0)++(vg) coordinate (aO0);
       \path (L0)++(vg) coordinate (aL0);
       \path (F0)++(vg) coordinate (aF0);

       \draw[<->,  opt] (aA0) -- (aO0) node[above=0.25cm, right=-0.9cm] {$\opt$};   
       \draw[<->,  lf] (aL0) -- (aF0) node[above=0.25cm, right=-0.55cm] {$\lf$}; 

       \draw (A1) -- (O1) -- (L1) -- (F1);
       \draw (A1) -- +(vn) node[below] {\footnotesize{$A_{k-1}$}};
       \draw (O1) -- +(vn); 
       \draw (L1) -- +(vn); 
       \draw (F1) -- +(vn) node[below] {$F_{k-1}$};;
       \path (A1)++(vg) coordinate (aA1);
       \path (O1)++(vg) coordinate (aO1);
       \path (L1)++(vg) coordinate (aL1);
       \path (F1)++(vg) coordinate (aF1);

       \draw[<->,  opt] (aA1) -- (aO1) node[above=0.25cm, right=-2cm] {$\opt$};   
       \draw[<->, lf] (aL1) -- (aF1) node[above=0.25cm, right=-1cm] {$\lf$}; 

       \draw (O2) -- (L2) -- (F2);
       \draw (L2) -- +(vn); 
       \draw (F2) -- +(vn) node[below] {$F_{k-2}$};;
       \path (L2)++(vg) coordinate (aL2);
       \path (F2)++(vg) coordinate (aF2);

       \draw[<->, lf] (aL2) -- (aF2) node[above=0.25cm, right=-2cm] {$\lf$}; 


       \draw[->,thin] (0, -6) -- (0, -2) node[right=1.5cm, above= -0.5cm] {Wait Ratio};
{\tiny
       \draw[<->,thin] (1, -6) -- (-24, -6);
       \foreach \y in {1, 2, 3} \draw (-0.1, \y - 6) -- (0.1, \y -6) node[right] {\y};

}

       \path (2,-5) coordinate (c10);
       \path (2,0) coordinate (c0);
       \path (2,0)++(vg1) coordinate (c1);
       \path (2,0)++(vg2) coordinate (c2);

       \draw (c0) -- (c0) node[right] {$\cj_k$};
        \draw (c1) -- (c1) node[right] {$\cj_{k-1}$};
        \draw (c2) -- (c2) node[right] {$\cj_{k-2}$};
        \draw (-11,-7) -- ++(0,0) node[below] {Time};

      \draw (-4, -6) -- (0, -6+2) node[above=0.2cm, left] {$\cj_k$};
      \draw (-17, -6) -- (-1, -6+ 1.5) node[above=0.2cm, left=0.5cm] {$\cj_{k-1}$};
      \draw (-24 , 0.8-6) -- (-5, 1.125-6) node[above = 0.15cm, left = 4cm] {$\cj_{k-2}$};

       \draw[<->, opt] (-4, -6) -- (-2, -6+1);
       \draw[<->, lf] (-1, -6+1.5) -- (0, -6+2);
       \draw[<->, opt] (-17, -6) -- (-9, -6+0.75);
       \draw[<->, lf] (-5, 1.125-6) -- (-1, -6+1.5);
       \draw[<->, lf]  (-21, 0.843 - 6) -- (-5, 1.125-6);

      \end{scope}


    \end{tikzpicture}
}

  \end{center}
  \vspace{-0.75cm}
  \caption{Comparison of scheduling of group $\cj_{k}$, $\cj_{k-1}$, and $\cj_{k-2}$ by $\lf$ and $\opt$.}
\end{figure}
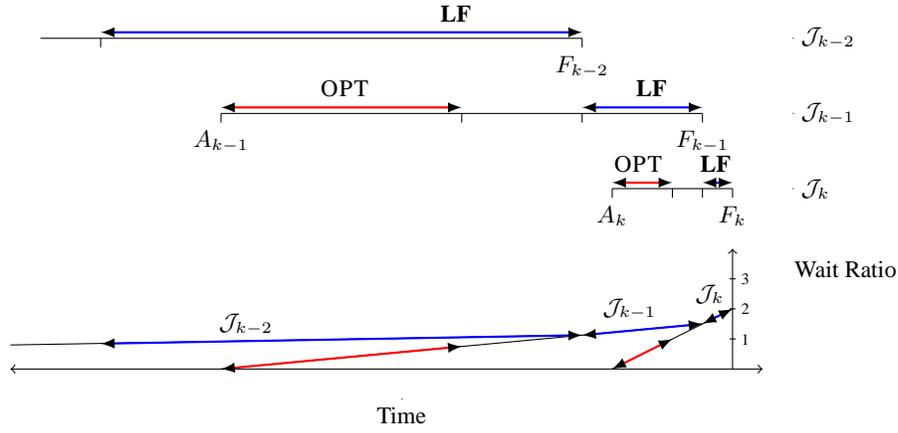

We now formally prove that $\lf$ achieves wait ratio $c$, while $\opt$ has wait ratio at most 1 for the given problem
instance $\sigma$. Let $F_i = A_i + (sc)^{k-i+1}, 0 \leq i \leq k$. Let $R_i$ be the maximum wait ratio for any job in
$\cj_i$ witnessed by $\lf$. We now define $k$ to be a constant such that $(1 - \frac{1}{sc})^k c \leq \frac{1}{3s}$.

\begin{lemma}
    \label{lem:ld_act}
    $\lf$, given speed $s$, processes $\cj_0$ during $[A_0, F_0]$ and
    $\cj_i$ during $[F_{i-1}, F_i]$, $1 \leq i \leq k$.
\end{lemma}
\begin{proof}[Proof Sketch:]
  By simple algebra one can check that the length of the time
  intervals $[A_0, F_0]$ and $[F_{i-1}, F_i]$ is the exact amount of
  time for $\lf$ with $s$-speed needs to completely process $\cj_0$
  and $\cj_i$, respectively.

  First we show that $\cj_0$ is finished during $[A_0, F_0]$ by $\lf$.
  It can be seen that at time $F_0$ the jobs in $\cj_j$ for $2 \leq j
  \leq k$ have not arrived, so we can focus on the class $\cj_1$.  The
  jobs in $\cj_1$ can be shown to have the same wait ratio as the jobs
  in $\cj_0$ at time $F_0$ and therefore the jobs in $\cj_1$ have
  smaller wait ratio than the jobs in $\cj_0$ at all times before
  $F_0$. This is because $\cj_0$ has a bigger slack than $\cj_1$.
  Hence, $\lf$ will finish all of the jobs in $\cj_0$ before beginning
  the jobs in $\cj_1$.

  To complete the proof, we show that $\cj_i$ is finished during
  $[F_{i-1}, F_i]$ by $\lf$.  It can be seen that at time $F_i$ the
  jobs in $\cj_j$ for $i+2 \leq j \leq k$ have not arrived, so we can
  focus on the class $\cj_{i+1}$.  The jobs in $\cj_i$ can be shown to
  have the same wait ratio as the jobs in $\cj_{i+1}$ at time $F_i$
  and therefore the jobs in $\cj_{i+1}$ have smaller wait ratio than
  the jobs in $\cj_i$ at all times before $F_i$.  Hence, $\lf$ will
  finish all of the jobs in $\cj_i$ before beginning the jobs in
  $\cj_{i+1}$.

\end{proof}

Using Lemma~\ref{lem:ld_act} and the given arrival times of each of
the jobs we have the following lemma.

\begin{lemma}
    \label{lem:ld_poor}
    $R_i = c(1 - \frac{1}{sc})^{k-i}$ ~for~ $0 \leq i \leq k$.
    \end{lemma}

    Notice that Lemma~\ref{lem:ld_poor} implies that $R_k \geq c$.
    Hence, the maximum delay factor witnessed by $\lf $ is at least
    $c$. In the following lemma, we show that there exists a valid
    scheduling by $\opt$ where the maximum wait ratio is at most one.
    This will show that $\lf$ achieves a competitive ratio of $c$.
    Note that Lemma~\ref{lem:ld_poor} shows that $R_0 \leq
    \frac{1}{3s}$.

\begin{lemma}
    \label{lem:opt_good}
    Consider a schedule which processes each job in $\cj_{0}$ during
    $[F_k, F_k + |\cj_{0}|]$ and each job in $\cj_{i}$ during $[A_i,
    A_i + |\cj_{i}|]$ for $1 \leq i \leq k$.  This schedule is valid
    and, moreover, the maximum wait ratio witnessed by this schedule
    is at most one.
\end{lemma}
\begin{proof}[Proof Sketch:]
  It is not hard to show that the time intervals $[F_k, F_k +
  |\cj_{0}|]$ and $[A_i, A_i + |\cj_{i}|]$ for $1 \leq i \leq k$ do
  not overlap, therefore this is a valid schedule.

  The wait ratio witnessed by the jobs in groups $\cj_i$ for $1 \leq i
  \leq k$ can easily be seen to be at most 1.  This is because this
  schedule processes each of these jobs as soon as they arrive.  We
  now show that the wait ratio of each of the jobs in $\cj_0$ is at
  most $1$. Recall that $R_0$, the maximum wait ratio of $\cj_0$ by
  $\lf$ is at most $\frac{1}{3s}$ at time $F_0$. Using the fact $sc
  \geq 2$, we can easily show that $|F_k - A_0| \leq 2 |F_0 - A_0|$.
  Note that $\opt$ can finish $\cj_0$ during $[F_k, F_k + s|F_0 -
  A_0|]$, since $\lf$ with $s$-speed could finish $\cj_0$ during
  $[A_0, F_0]$. Thus the wait ratio of $\cj_0$ at time $F_k + s|F_0 -
  A_0|$ is at most $\frac{2+s}{3s} \leq 1$.
\end{proof}

From Lemma~\ref{lem:ld_poor} the maximum delay factor witnessed by
$\lf$ is $c$ and by Lemma~\ref{lem:opt_good} the maximum delay factor
witnessed by $\opt$ is $1$. Hence we have the proof of
Theorem~\ref{thm:lf}.

\section{Conclusion}
In this paper, we showed an almost fully scalable
algorithm\footnote{An algorithm is said to be almost fully scalable if
  for any fixed $\eps>0$, it is $O(1+\eps)$-speed
  $O(1)$-competitive.} for minimizing the maximum delay factor in
broadcasting for unit sized jobs. The slight modification we make to
$\sbg$ from \cite{ChekuriM09} makes the algorithm more practical. Using the
intuition developed for the maximum delay factor, we proved that
$\fifo$ is in fact 2-competitive for varying sized jobs closing the
problem for minimizing the maximum response time online in broadcast
scheduling.

We close this paper with the following open problems. Although the new
algorithm for the maximum delay factor with unit sized jobs is almost
fully scalable, it explicitly depends on speed given to the
algorithm. Can one get another algorithm independent of this
dependency? For different sized pages, it is still open on whether
there exists a $(1+\epsilon)$-speed algorithm that is
$O(1)$-competitive.  For minimizing the maximum response time offline
it is of theoretical interest to show a lower bound on the
approximation ratio that can be achieved or to show an algorithm that
is a $c$-approximation for some $c < 2$.

\bibliographystyle{plain}
\bibliography{MDF}

\appendix

\section{Omitted Proofs}
\subsection{Proof of Lemma~\ref{lem:main}}
\begin{proof}
Let $J_{x,i}, J_{x,j} \in \cji$ such that $i < j$. Let $t'$ be the time that $\sbg$ starts satisfying request
$J_{x,i}$. By the definition of $I$, request $J_{x,i}$ must have delay factor at least $\alpha^*$ at time $f_{x,i}$.
We also know that the request $J_{x,j}$ must arrive after time $t'$, otherwise request $J_{x,j}$ must also be satisfied
at time $t'$.  If the optimal solution combines these requests into a single broadcast then the request $J_{x,i}$ must
wait until the request $J_{x,j}$ arrives to be satisfied. However, this means that the request $J_{x,i}$ must achieve a
delay factor greater than $\alpha^*$ by $\opt$, a contradiction of the definition of $\alpha^*$.
\end{proof}

\subsection{Proof of Lemma~\ref{lem:len-varying}}

\begin{proof}
 The request $J_{q,k}$ has delay factor at least $c\alpha^*$ at any time $t$ during $I' = [t', t^*]$, where $t' = t^* -(c^2
-c)\st_{q,k}\alpha^*$. The largest delay factor any request can have during $I'$ is less than $c^2 \alpha^*$ by
definition of $t^*$ being the first time $\sbg$ witnesses delay factor $c^2 \alpha^*$. Hence the request $J_{q,k}$ is
in the queue $Q(t)$ at any time $t$ during $I'$. Therefore, any request that forced $\sbg$ to broadcast on $I'$, must
have delay factor at least $\alpha^*$ and since $J_{q,k} \in Q(t)$ for all $t \in I'$, the requests scheduled on $I'$ must have slack at
most $\st_{q,k}$.
\end{proof}

\subsection{Proof of Lemma~\ref{lem:arrivaltime-varying}}

\begin{proof}
For the sake of contradiction, suppose that some request $J_{p,i}$ that forced $\sbg$ to broadcast page $p$ on the
interval $I$ arrived at time  $t' < t_1 -2 \st_{q,k}  \alpha^* c$. Recall that $J_{p,i}$ has a slack no bigger than
$S_{q,k}$ by the definition of $I$. Therefore at time $t_1 - \st_{q,k} \alpha^* c$, $J_{p,i}$ has a delay factor of at
least $c \alpha^*$. Thus any request scheduled during the interval $[t_1 - \st_{q,k} \alpha^* c, t_1]$ has a delay
factor no less than $\alpha^*$. We observe that $J_{p,i}$ is in $Q(\tau)$ for $\tau \in [t_1 - \st_{q,k} \alpha^* c,
t_1]$; otherwise there must be a request with a delay factor bigger than $c^2 \alpha^*$ at time $\tau$ and this is
a contradiction to the assumption that $t^*$ is the first time that $\sbg$ witnessed a delay factor of $c^2 \alpha^*$.
Therefore any request that forced $\sbg$ to  broadcast during $[t_1 - \st_{q,k} \alpha^* c, t_1]$ has a slack
no bigger than $S_{p,i}$. Also we know that $S_{p,i} \leq S_{q,k}$ by the definition of $I$. In sum, we showed that any
request that forced $\sbg$ to do a broadcast during $[t_1 -  \st_{q,k} \alpha^* c, t_1]$ have a slack no bigger than
$S_{q,k}$ and a delay factor no smaller than $\alpha^*$, which is a contradiction of the definition of $t_1$.
\end{proof}

\end{document}